\documentclass[conference]{IEEEtran}
\IEEEoverridecommandlockouts
\usepackage{cite}
\usepackage{amsmath,amssymb,amsfonts}
\usepackage{xcolor}
\usepackage[
    colorlinks=true,
    linkcolor=blue,
    citecolor=blue,
    urlcolor=blue
]{hyperref}

\usepackage{algorithmic}
\usepackage{graphicx}
\usepackage{textcomp}

\usepackage{subfig}
\usepackage{siunitx}
\newcommand{\ours}{\text{MSNF}}

\def\BibTeX{{\rm B\kern-.05em{\sc i\kern-.025em b}\kern-.08em
    T\kern-.1667em\lower.7ex\hbox{E}\kern-.125emX}}
\begin{document}

\title{Masked Wavelet Scattering Transform Neural Field for Sound Field Reconstruction\\

\thanks{This research was funded by the Fonds de recherche du Québec – Nature et technologies (FRQNT) Doctoral Research Scholarship [363876, \href{https://doi.org/10.69777/363876}{DOI: 10.69777/363876}}
}


\author{
\IEEEauthorblockN{Xinmeng Luan}
\IEEEauthorblockA{
\textit{Schulich School of Music,} \\
\textit{McGill University, } \\
\textit{CIRMMT$^\dagger$\thanks{$^\dagger$Centre for Interdisciplinary Research in Music Media and Technology (CIRMMT)} } \\
Montreal, Canada \\
xinmeng.luan@mail.mcgill.ca, 
}
\and
\IEEEauthorblockN{Samuel A. Verburg}
\IEEEauthorblockA{
\textit{Dept. of Electrical and } \\
\textit{Photonics Engineering, }\\
\textit{Technical University of Denmark} \\
Kgs. Lyngby, Denmark \\
saveri@dtu.dk
}
\and
\IEEEauthorblockN{Efren Fernandez-Grande}
\IEEEauthorblockA{
\textit{Dept. of Audiovisual  } \\
\textit{Engineering and }\\
\textit{ Communications, Universidad}\\
\textit{ Polit\'ecnica de Madrid} \\
Madrid, Spain \\
efren.fernandez@upm.es
}
\and
\IEEEauthorblockN{Gary Scavone}
\IEEEauthorblockA{
\textit{Schulich School of Music,} \\
\textit{McGill University, } \\
\textit{CIRMMT$^\dagger$} \\
Montreal, Canada \\
gary.scavone@mcgill.ca
}
}



\maketitle

\begin{abstract}
In this paper, we propose a reconstruction framework that leverages the Wavelet Scattering Transform (WST) as a multi-scale feature extractor to impose statistical priors under sparse observation conditions. The reconstruction problem is formulated as an optimization task and solved using a neural field, with the WST incorporated into the training loss function. As a proof of concept, we validate the proposed method on HRTF upsampling. A masking strategy is applied to the WST coefficients, resulting in a two-phase procedure. The first phase learns a binary mask from a small multi-subject dataset, while the second phase applies the learned mask to the WST coefficients of an individual HRTF to preserve informative statistical structures during reconstruction. Validation against baseline methods, which also serve as an ablation study of the different components of the framework, demonstrates the effectiveness of the proposed approach.

\end{abstract}

\vspace{-10pt} 
\section{Introduction}
\label{sec:introduction}

Sound field reconstruction refers to the task of recovering a sound field from a limited set of  sparse measurements, which are often obtained using microphone arrays \cite{williams1999fourier}. In a spherical setting, this problem encompasses tasks such as Head-Related Transfer Function (HRTF) upsampling \cite{hogg2025listener}. 
Due to the large experimental effort involved in measuring HRTFs, available datasets remain small, especially compared with the amount of data accessible in computer vision and other fields. 
The challenge is compounded by the fact that HRTFs are highly individualized, as they are determined by each listener’s unique anatomical features. While HRTFs across individuals may share certain similarities at specific scales, they can differ substantially at others. Deep learning approaches trained on large datasets attempt to implicitly learn these shared structures; however, due to limited data availability, they are prone to overfitting \cite{zhang2023hrtf}.
To address this issue, we propose an HRTF upsampling method that explicitly leverages multi-scale representations to capture and preserve relevant structures in a controllable manner, even under limited data conditions, ensuring that only the features exhibiting consistent behavior across the limited available dataset are retained.

The Wavelet Scattering Transform (WST) \cite{mallat2012group, bruna2013invariant} appears to be a promising candidate for this task. It extracts multi-scale features through a cascade of wavelet convolutions and nonlinear modulus operators, producing a translation-invariant representation of the signal. The WST can be interpreted as a convolutional neural network (CNN) with fixed, analytically defined filters rather than learned ones \cite{mallat2012group, bruna2013invariant}. Unlike conventional CNNs, the WST is highly interpretable and is supported by well-defined mathematical guarantees \cite{mallat2012group, bruna2013invariant}.
Studies have shown that the WST outperforms end-to-end learned CNNs in scenarios with scarce datasets \cite{oyallon2018scattering}.
The WST has been explored across a variety of research fields, including images \cite{bruna2013invariant, oyallon2018scattering, bruna2019multiscale}, physics fields \cite{cheng2024scattering, mousset2024generative}, audio signals \cite{anden2015joint, han2024learning, wang2022adaptive, lostanlen2019shape} and financial signals \cite{morel2025scale}, among others.

In this paper, we propose a reconstruction framework that leverages the WST as a multi-scale feature extractor to provide \textit{statistical priors} under sparse observation conditions. The reconstruction problem is formulated as an optimization task and solved using a neural field (NF) representation, which models the target as a continuous field. As a proof of concept, we validate the proposed approach on HRTF upsampling.
The proposed method introduces a masking strategy applied to the WST coefficients, resulting in a two-phase procedure. In the first phase, referred to as \textit{mask identification}, a binary mask is learned from a small dataset comprising multiple subjects. In the second phase, referred to as \textit{neural field reconstruction}, the mask is applied to the WST coefficients of an individual HRTF to preserve the informative statistical features during reconstruction. During training, the loss function combines a data fidelity term defined on the sparse observations with a regularization term based on the masked WST coefficients.
We term the resulting framework the Masked Wavelet Scattering Transform Neural Field (\ours{}).

To the best of our knowledge, this is the first work to exploit the WST as a prior for sound field reconstruction. More broadly, the proposed framework is not restricted to HRTFs or even to acoustic data. It can be applied to a wide range of reconstruction and synthesis problems in data-scarce settings. In addition, the formulation is highly flexible and can be readily adapted to a wide range of engineering applications.

\vspace{-3pt} 
\section{Related Work}
\label{sec:background}
\vspace{-5pt} 

\subsection{Sound Field Reconstruction}
\label{subsec:SoundFieldReconstruction}

\subsubsection{Problem formulation}

Consider a sound field in the frequency domain, represented by the sound pressure signal $p(u) \in \mathbb{R},  u = (u_1, u_2)$, where $(u_1, u_2)$ specifies the spatial position in a $2$-dimensional space.  In this simplified setting, only the magnitude component of the sound pressure is considered.
$p(u)$ is defined over a continuous spatial domain and represents the idealized sound field at any point in space. In practice, only a finite set of measurements can be obtained using microphone arrays, yielding a discretely sampled version of the field denoted by $p[u]$.

The sound field reconstruction problem aims to recover the full sound field from a set of sparsely sampled measurements $p[u] \in \mathbb{R}^2$, typically obtained using microphone arrays. The objective is to reconstruct the sound field across the entire spatial domain, ideally in continuous form, though in practice a sufficiently high-resolution approximation is usually adequate. This task is also commonly referred to as an upsampling problem.

\subsubsection{Deep learning application}
Deep learning-based methods have become increasingly popular for addressing this inverse problem \cite{fernandez2023generative, olivieri2024physics, verburg2025differentiable,  olivieri2021physics, luan2025pisfd}.
In the context of HRTF interpolation, neural field (NF)-based approaches have recently received increasing attention \cite{zhang2023hrtf, masuyama2025retrieval, masuyama2025sudafield}. 

\vspace{-5pt} 
\subsection{Wavelet Scattering Transform}
\label{subsec:WaveletScatteringTransform}

\subsubsection{Theory}
The following description is based on the formulations in \cite{bruna2013invariant, oyallon2018scattering}.
Let $S^{i}$ denote the scattering operator of order $i$, where $S^{i} p[u]$ represents the $i$-th order scattering coefficients. Let $\phi_J$ be a local averaging filter with spatial support at scale $2^J$, where $J \in \mathbb{N}$ is the corresponding scale parameter.
The zeroth-order scattering coefficients are 
\begin{equation}
    S^0 p[u] = p *\phi_J (2^J u).
\end{equation}
Here $*$ denotes convolution.
The zeroth-order scattering transform is approximately invariant to translations smaller than $2^J$, but also results in a loss of high frequencies, which are recovered by using higher order scattering transform with wavelets \cite{oyallon2018scattering}.
A wavelet is an integrable function with zero mean, localized both in Fourier and space domain. 
A family of wavelets is obtained by dilating and rotating a complex mother wavelet $\psi$ (here, a Morlet wavelet) such that $\displaystyle{\psi_{j, \theta} = \frac{1}{2^{2j}} \psi(r_{-\theta }\frac{u}{2^j})} $, where $r_{-\theta}$ is the rotation by $-\theta$, and $j \geq 0$ is the scale of the wavelet. 
Let $L \in \mathbb{N}$ denote a number of discrete angles uniformly sampling the interval $[0,2\pi]$.
A wavelet transform $W_1$ computes convolutions of a signal  with the family of wavelets
\begin{equation}
    W_1p[j_1, \theta_1, u] = \{ p* \psi_{j_1, \theta_1}(2^{j_1} u )\}_{j_1 \leq J, \theta_1 = 2 \pi \frac{l}{L}, 1 \leq l \leq L} \ .
\end{equation}
To obtain non-trivial translation invariance, a nonlinear operator is required \cite{bruna2013invariant}. Here, a pointwise complex modulus is utilized. Thus, the first-order WST coefficients are
\begin{equation}
    S^1p[j_1, \theta_1, u] = |p* \psi_{j_1, \theta_1}|* \phi_J (2^J u).
\end{equation}
We apply a second wavelet transform $W_2$, using the same filters as $W_1$, to the first-order WST coefficients before the averaging step. This recovers the high-frequency information lost due to the averaging applied on the first-order coefficients \cite{oyallon2018scattering} and leads to the second-order WST coefficients 
\begin{equation}
\begin{aligned}
    S^2p[j_1,j_2,\theta_1,\theta_2,u]  &=||  p* \psi_{j_1, \theta_1}| * \psi_{j_2, \theta_2}|* \phi_J (2^J u). 
    \end{aligned}
\end{equation}
Note that we only compute paths of increasing scale $(j_1 < j_2)$ since non-increasing paths have been shown to bear no energy \cite{bruna2013invariant}. The scattering expansion is truncated at second order, since higher-order scattering coefficients typically contribute very little energy and therefore are considered negligible in practice \cite{bruna2013invariant}. The final scattering coefficient corresponding to the concatenation of the order 0, 1 and 2 WST coefficients is denoted as 
\begin{equation}
    Sp[u] = \{S^0p[u], S^1p[j_1, \theta_1, u], S^2p[j_1, j_2, \theta_1, \theta_2, u] \}.
\end{equation}
There are in total $N_{co}= 1 + JL + \frac{1}{2} J(J-1) L^2$ WST coefficients.

\subsubsection{Application}
The WST was originally proposed and demonstrated to
be successful in classification tasks \cite{mallat2012group, bruna2013invariant, oyallon2018scattering}, and has since
been extended to data generation and synthesis applications \cite{bruna2019multiscale, cheng2024scattering, mousset2024generative, morel2025scale}.
Among these approaches, the Microcanonical Gradient Descent model \cite{bruna2019multiscale} has recently gained attention in cosmology for the generation of random fields \cite{cheng2024scattering, mousset2024generative}. This framework typically assumes that the underlying random field is statistically homogeneous (or stationary) \cite{cheng2024scattering, mousset2024generative, morel2025scale}, meaning that its joint probability distributions are invariant under translations \cite{albeverio1975homogeneous}.
The advantage of using the WST in this context is its ability to capture multi-scale statistics, which outperforms generating Gaussian random fields that only preserve up to the second-order moments \cite{cheng2024scattering,morel2025scale}.

In audio applications, the WST has been employed for audio classification \cite{anden2015joint}, playing technique recognition \cite{wang2022adaptive}, audio texture synthesis  \cite{lostanlen2019shape}
and sound matching \cite{han2024learning}. 

The WST is commonly used either as a backbone model (effectively replacing CNNs and requiring no trainable neural network components) \cite{cheng2024scattering, mousset2024generative, oyallon2018scattering}, or as a fixed encoder combined with learnable neural networks \cite{angles2018generative}.
In \cite{han2024learning}, Han et al. employ the Joint Time-Frequency Scattering transform (JTFS) to extract audio features and incorporate them into the loss function of an optimization framework for perceptual sound matching. This usage shares a similar intuition with the approach proposed in this work.

\vspace{-5pt} 
\section{Proposed Method}
\label{sec:method}
\vspace{-2pt}

\begin{figure}
    \centering
    \includegraphics[width=1\linewidth]{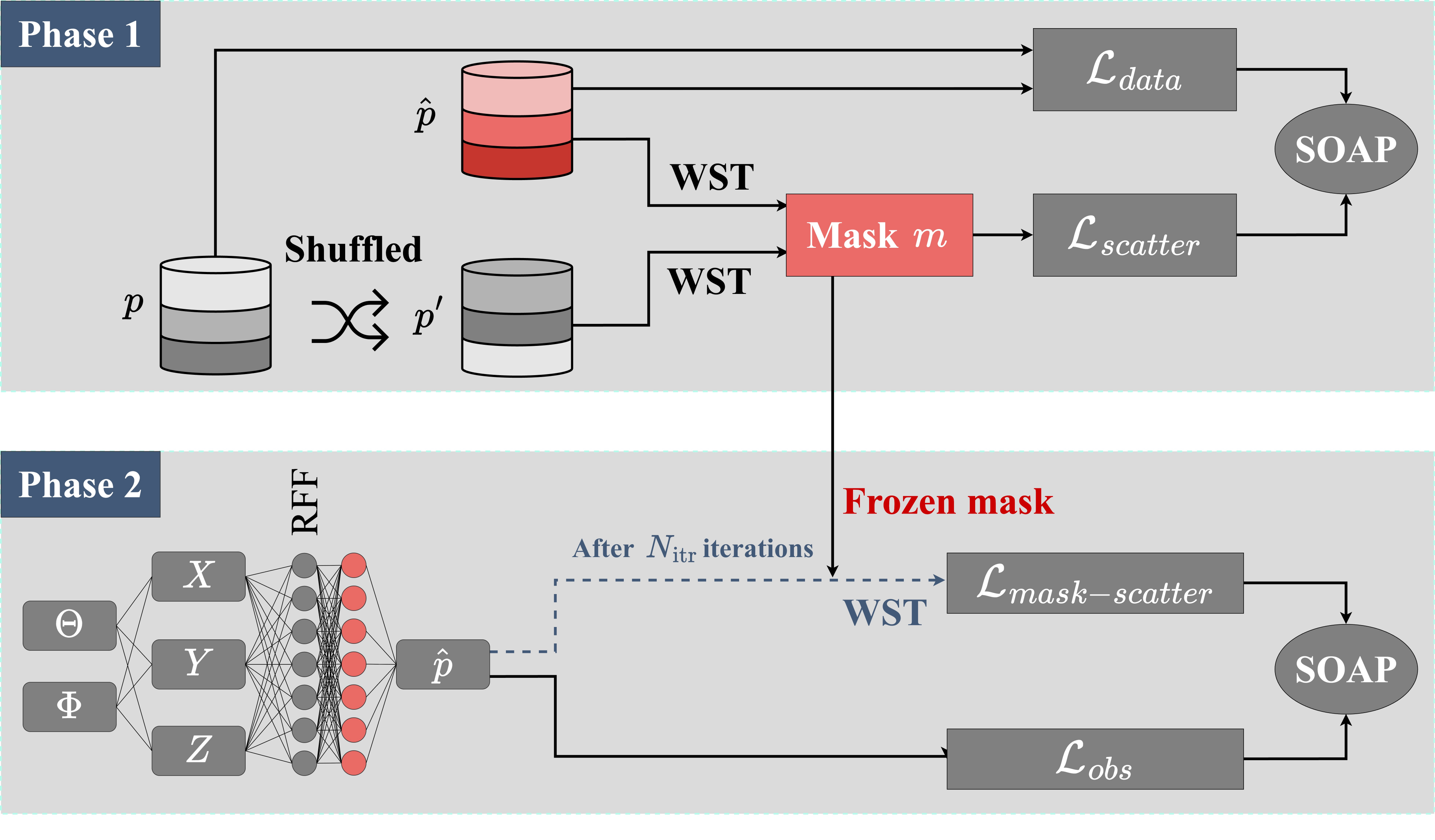}
    \caption{The proposed \ours{}  framework. Top: Phase~1, mask identification; Bottom: Phase~2, neural field reconstruction. Components shown in red correspond to trainable parameters.}
    \label{fig:config}
    \vspace{-16pt} 
\end{figure}

We propose the Masked Wavelet Scattering Transform Neural Field (\ours), which consists of two phases, as shown in Fig.~\ref{fig:config}. In Phase~1, mask coefficients are learned from a limited dataset, a process referred to as \textit{mask identification}. In Phase~2, an MLP-based neural field backbone is employed to reconstruct the signal in the continuous domain, informed by the masked scattering coefficients and the observed sparse data. We refer to this stage as \textit{neural field reconstruction}. The trainable quantities being optimized are highlighted in red in Fig.~\ref{fig:config}.

\subsection{Phase~1: Mask Identification}
\vspace{-5pt} 
An approach using different masks in the pixel space to handle non-homogeneous field modeling with the WST was proposed in \cite{delouis2022non}. However, in that work, the masks are manually defined. To overcome this limitation, we aim to develop an adaptive method for mask determination. Given $M$ target data realizations, our goal is to learn a weighting mask $m \in \mathbb{R}^{N_{\mathrm{co}}}$, that assigns larger values to the WST coefficients showing strong similarity across the realizations, and smaller values otherwise. The mask is applied to the WST coefficients via Hadamard product, as $ m\odot Sp[u]$.

This task is formulated as a joint optimization problem, where both the 
$M$ predicted data realizations $\hat{p} [u_1, u_2] $ and the mask $m$ are learned simultaneously. 
The first term in Eq.~\eqref{eq:loss} is the data loss, defined as the Mean Squared Error (MSE) between the $k$-th predicted and target data. The second term is the scattering loss, defined as the MSE of the masked scattering coefficients of the $k$-th predicted data with a randomly selected target data $p'[u_1, u_2]$ from the dataset. Without this term, the task would be trivial. The scattering loss is introduced to enforce similarity of scattering coefficients, by minimizing the difference between the masked scattering coefficients for predicted and randomly chosen target data.
The overall loss function is defined as
\begin{equation}
\begin{aligned}
&\mathcal{L}_{\text{data}} = \frac{1}{M} \frac{1}{N_1 N_2} \sum_{k=1}^{M} \sum_{i=1}^{N_1} \sum_{j=1}^{N_2} \left( \hat{p}_k[u_{1_i}, u_{2_j}] - p_k[u_{1_i}, u_{2_j}] \right)^2, \\&
\mathcal{L}_{\text{scatter}} = \frac{1}{M} \frac{1}{N_{\text{co}}} \sum_{k=1}^{M} \sum_{i=1}^{N_{\text{co}}} \left( m_i \big( S\hat{p}_k[u_1, u_2]_i - Sp'_{k}[u_1, u_2]_i  \big) \right)^2, \\&
\mathcal{L}_1 = \mathcal{L}_{\text{scatter}} + \alpha \mathcal{L}_{\text{data}},
\label{eq:loss}
\end{aligned}
\end{equation}
where $\alpha$ is a weighting factor, chosen as a small positive parameter, and $N_1$ and $N_2$ denote the numbers of spatial points associated with  $u_1$ and $u_2$, respectively.
Batch learning is employed, with each batch containing 
$M$ distinct target data realizations, along with randomly selected scattering coefficients from all target data.

\vspace{-5pt} 
\subsection{Phase~2: Neural Field Reconstruction}
\vspace{-5pt} 
Once the mask is obtained from Phase~1, the masked scattering coefficients can be used as a prior to guide the learning process in the reconstruction.
An MLP is adopted as the backbone to represent the continuous field, referred to as a neural field. The MLP takes spatial coordinates $(u_1, u_2)$ as input and outputs the corresponding acoustic pressure $p$. 
Moreover, the Random Fourier Feature (RFF) embedding \cite{tancik2020fourier} is applied to the input data, mapping it into a higher-dimensional space through sinusoidal transformations. This helps alleviate the spectral bias phenomenon, namely the tendency of neural networks to learn low-frequency components more quickly while struggling to capture high-frequency ones \cite{rahaman2019spectral}.
Let the neural field be denoted by
\begin{equation}
    p = \mathcal{F}_\gamma (u_1, u_2),
\end{equation}
where $\mathcal{F}_\gamma$ represents the network and  $\gamma$ represents the set of trainable parameters.

The loss function consists of two terms. The first is the observation loss, defined as the MSE between the predicted and observed data at the sparse measurement locations,
\begin{equation}
    \mathcal{L}_{\text{obs}} = \frac{1}{N^2}  \sum_{i=1}^{N_1} \sum_{j=1}^{N_2} \left( \hat{p}[u_{1_i}, u_{2_j}] - p[u_{1_i}, u_{2_j}] \right)^2.
    \label{eq:ob}
\end{equation}
The second term is the masked scattering coefficient loss, defined as the MSE between the predicted data and a selected target realization after applying the WST at the corresponding spatial locations,
\begin{equation}
\mathcal{L}_{\text{mask-scatter}} =  \frac{1}{N_{co}} \sum_{i=1}^{N_{co}} \left( m_i \big( S\hat{p}[u_1, u_2]_i - Sp'[u_1, u_2]_i  \big) \right)^2.
\label{eq:mask_scatter}
\end{equation}
The total loss is 
\begin{equation}
    \mathcal{L}_2 = \beta_1\mathcal{L}_{\text{obs}} + \beta_2\mathcal{L}_{\text{mask-scatter}},
    \label{eq:total2}
\end{equation}
where $\beta_1, \beta_2$ are the weighting factors. Essentially, the masked scattering coefficient loss imposes a statistical prior, serving as a form of regularization.
Separate datasets are used for the loss functions \eqref{eq:ob} and \eqref{eq:total2}, based on the available spatial coordinates, and full-batch training is applied.

Additionally, we employ a two-subphase training strategy: during the first $N_{\text{itr}}$ iterations, the network is trained solely using the observation loss \eqref{eq:ob}, after which the training proceeds with the total loss \eqref{eq:total2}. The GradNorm \cite{chen2018gradnorm} strategy is also applied to adaptively balance the weights of each component in the loss function, $\beta_1$ and $\beta_2$.
The reconstruction is carried out in a one-shot manner, requiring a separate network to be trained for each HRTF.

In summary, the trainable parameters in the two training phases are: in Phase 1, the mask $m$ and the estimated field $ \hat{p}[u_1, u_2] $.; in Phase 2, only $ \gamma $.


\vspace{-5pt} 
\section{Validation}
\label{sec:validation}

\vspace{-5pt} 
\begin{figure*}
\vspace{-12pt} 
    \centering
    \subfloat[Ground truth]{\includegraphics[width=0.15\linewidth]{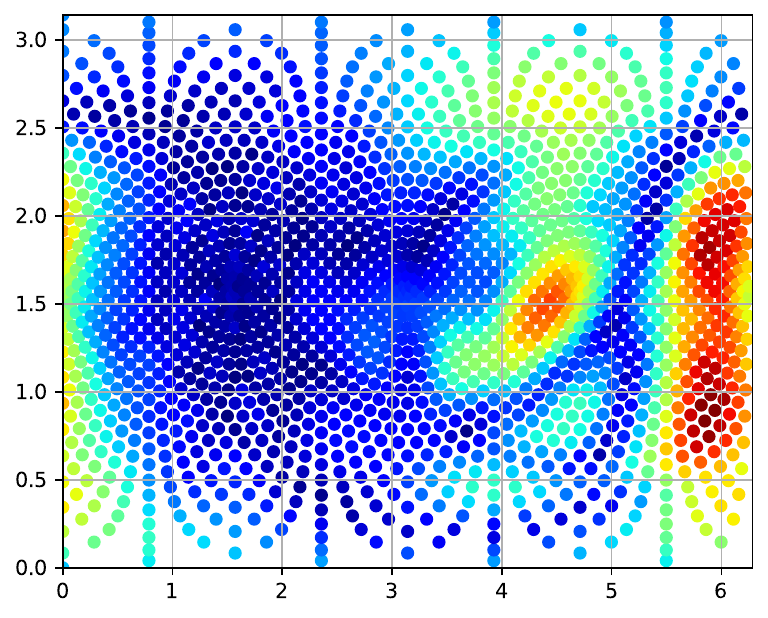}}
    \subfloat[\ours]{\includegraphics[width=0.15\linewidth]{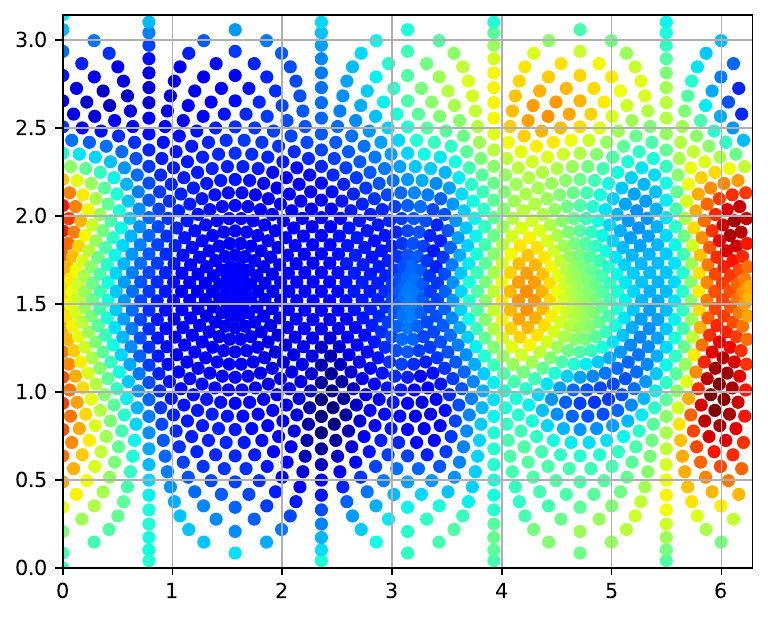}}
    \subfloat[SNF]
{\includegraphics[width=0.15\linewidth]{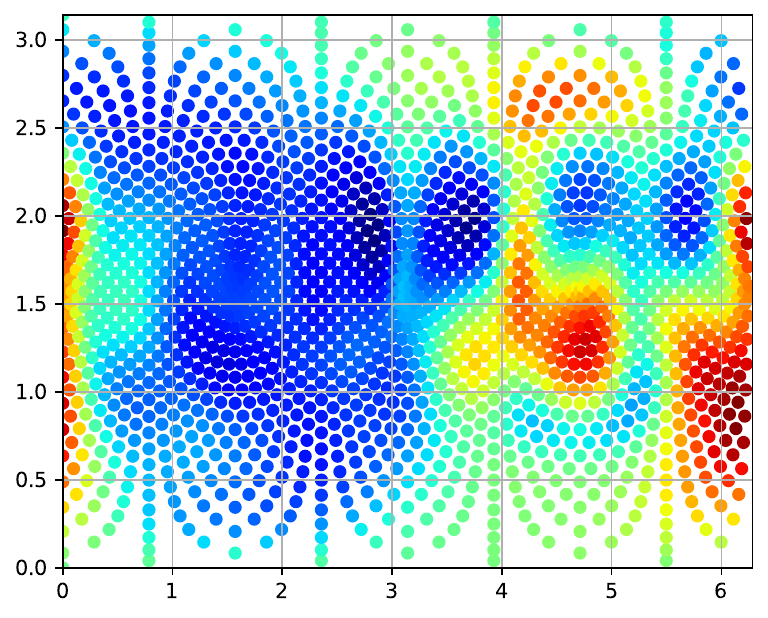}}
    \subfloat[NF]{\includegraphics[width=0.15\linewidth]{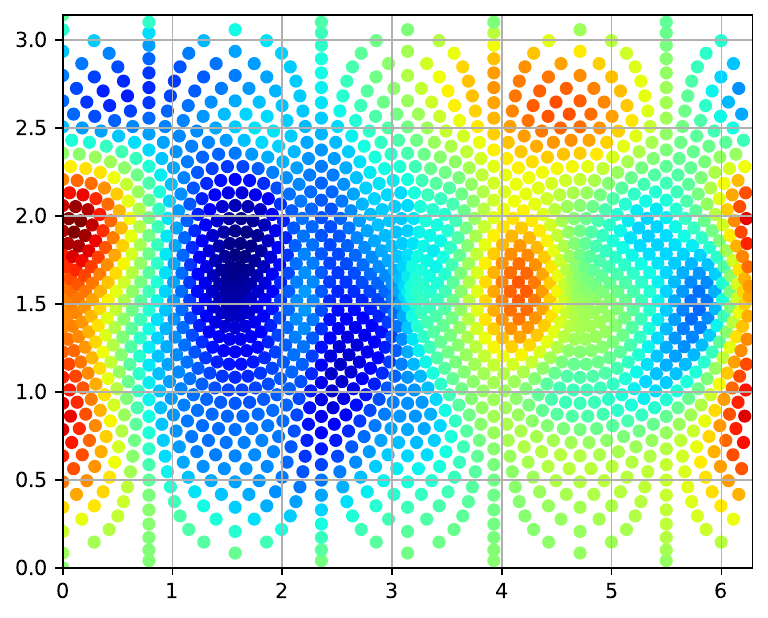}}
    \subfloat[SH]{\includegraphics[width=0.15\linewidth]
{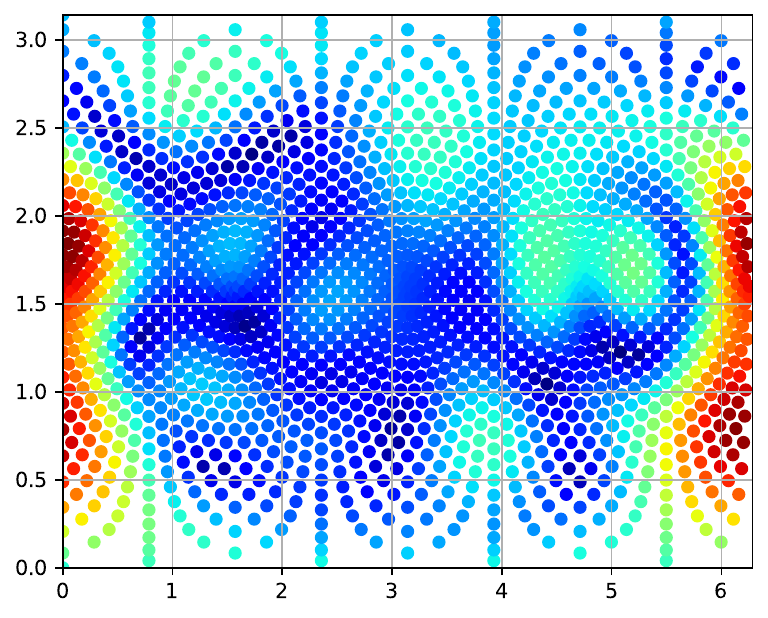}} 
    \subfloat[Reference sub. 11]{\includegraphics[width=0.15\linewidth]
{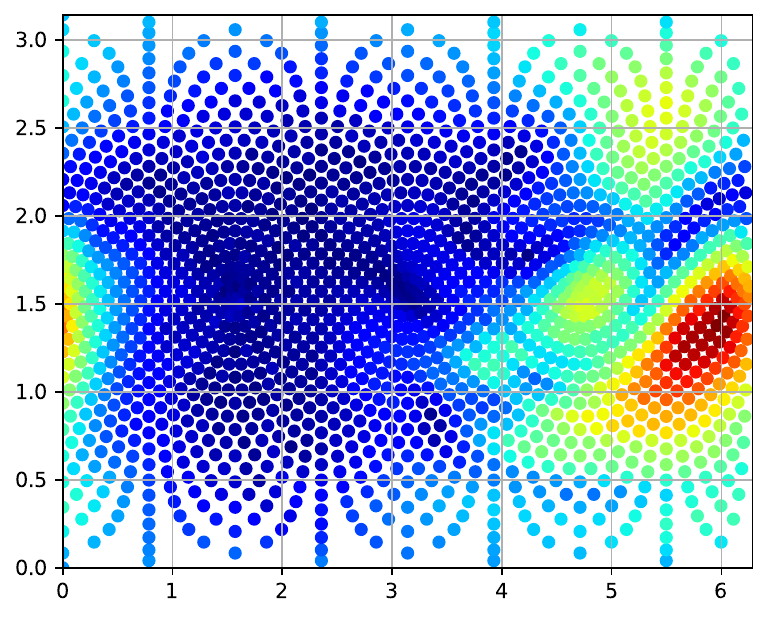} \label{fig: 1g}} 
    \subfloat{\includegraphics[width=0.025\linewidth]
    {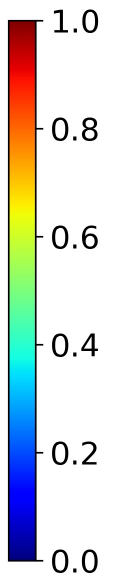}} 
    \caption{HRTF of subject 6 at \SI{14470}{\hertz}, right ear. The pressure is normalized to the range $[0, 1]$. The interpolation was performed from a $7 \times 7$ equally spaced observation grid to 1730 spatial points, matching the ground truth dataset \cite{brinkmann2019cross}. Axes: $x: \Phi$, $y: \Theta$.}
    \label{fig: example}
    \vspace{-15pt} 
\end{figure*}

\subsection{Dataset}
\vspace{-5pt} 
\sloppy
Experiments were conducted on the HRTF open source HUTUBS simulation dataset \cite{brinkmann2019cross}, which has a relatively high resolution with 1730 spatial angles. The dataset covers azimuth angles $\Theta \in [\SI{-60}{\degree}, \SI{-54}{\degree}, \ldots, \SI{60}{\degree}]$ and elevation angles $\Phi \in [\SI{4}{\degree}, \SI{10}{\degree}, \ldots, \SI{358}{\degree}]$, computed at a radius of \SI{0.74}{\meter}.
We evaluated the HRTF at $B=7$ frequencies, $[2067, 4134, 6200, 8269, 10336, 12403, 14470]~\si{\hertz}$, in alignment with \cite{ma2023spatial}.
A subset of $D=10$ subjects (Subjects 11–20) was used in Phase~1 to identify the mask. In Phase~2, we adopt the scattering coefficients of Subject~11 as the reference for the masked scattering coefficients loss~\eqref{eq:mask_scatter}, although any subject from Phase~1 could be selected.  The evaluation of different methods was then conducted on Subjects 6–10. The left and right ears were processed and evaluated independently across the dataset, treated as two separate channels.

\vspace{-5pt} 
\subsection{Comparison Approaches}
\vspace{-5pt} 
Three other approaches were evaluated to compare with \ours: Spherical Harmonic (SH) interpolation, Neural Field (NF), and Scattering Neural Field (SNF) (without mask). SH interpolation is a commonly used method for interpolating data defined on the sphere \cite{zotkin2009regularized, ahrens2020computation}. The HRTF can be decomposed using a truncated SH expansion, where the coefficients are estimated via a least-squares minimization procedure \cite{zotkin2009regularized}. Once obtained, these coefficients allow the HRTF to be reconstructed at any desired sampling scheme. In our experiment, the SH expansion was truncated at the 4th order, which was found to provide the best performance based on the experimental results (refer to \cite{zotkin2009regularized, ahrens2020computation} for more technical details). SH serves as a conventional baseline. 
Moreover, we tested the same NF architecture used in Phase~2, trained solely with the observation loss \eqref{eq:ob}. 
SNF refers to the version of \ours{} without the masking procedure. It only optimizes \eqref{eq:total2} and does not adopt the two-subphase training strategy.



\subsection{Implementation}
\vspace{-5pt}

Our implementation was carried out in \texttt{JAX}, with the \texttt{Kymatio} library \cite{andreux2020kymatio} employed for the WST. Each channel, frequency, and subject HRTF was treated as a 2D image, where the coordinates $(u_1, u_2)$ correspond to $(\Theta, \Phi)$. 
While formulating the problem in spherical coordinates might lead to improve performance, we found that treating the HRFT as images was sufficient to introduce the proposed general framework. 
Moreover, we used the magnitude of the HRTF data for simplicity, while the real and imaginary parts could alternatively be treated as two separate channels.

Since the \texttt{Kymatio} implementation in the 2D case requires an equally sampled grid, we resampled the data onto a $24 \times 24$ ($N_1=N_2=24$) regular grid with $\Theta \in [0 , \pi]$ and $\Phi \in [0, 2\pi]$ for the WST. This resampling is performed using SH according to \cite{ahrens2020computation}. 
Similarly, the observation data for testing was resampled to a $7 \times 7$ grid, using SH \cite{ahrens2020computation}. In both cases, the spherical harmonics are truncated at the 35th order, consistent with \cite{ma2023spatial}.

The batch size in Phase~1 was set to 10, with $\alpha = 10^{-3}$. The MLP receives $(\Theta, \Phi)$ as input, which is first converted to Cartesian coordinates $(X, Y, Z)$, then processed through an RFF layer with a scale parameter of 1, followed by a single hidden layer of 256 units with a $\tanh$ activation function.
Phase~1 concluded at epoch 200. The neural networks were trained for a total of 400 epochs, whereas for \ours{}, the second subphase in Phase~2 began at epoch 100 ($N_{itr}$ = 100).
We employ the
state-of-the-art second order optimizer ShampoO with Adam in
the Preconditioner’s eigenbasis (SOAP) \cite{vyassoap}.

\vspace{-5pt} 
\subsection{Metrics}
\vspace{-5pt} 
Three metrics were used for evaluation: Log-Spectral Distortion (LSD), Normalized Mean Square Error (NMSE), and Normalized Cross-Correlation (NCC). 
LSD quantifies the spectral similarity between a reference HRTF and an upsampled estimate across directions and frequency bins, defined as
\begin{equation}
    \text{LSD} = \frac{1}{N_1 N_2} \sum_{i=1}^{N_1}  \sum_{j=1}^{N_2} \sqrt{\frac{1}{B} \sum_{b=1}^B\left ( 20 \log_{10} \frac{|{{p}[u_{1_i},u_{2_j}]}_b|}{|{\hat{p}[u_{1_i},u_{2_j}]}_b|}\right )^2}.
\end{equation}
The NMSE is given by
\begin{equation}
\text{NMSE} = \frac{\sum_{i=1}^{N_1} \sum_{j=1}^{N_2} ({p}[u_{1_i},u_{2_j}] - \hat{p}[u_{1_i}, u_{2_j}] )^2}{\sum_{i=1}^{N_1}\sum_{j=1}^{N_2} {p}[u_{1_i},u_{2_j}]^2},
\end{equation}
and the NCC is given by
{\scriptsize
\begin{equation}
\text{NCC} = {\frac{\sum_{i=1}^{N_1} \sum_{j=1}^{N_2} ({p}[u_{1_i},u_{2_j}] - \bar{p})(\hat{p}[u_{1_i},u_{2_j}] - \bar{\hat{p}})}{\sqrt{\sum_{i=1}^{N_1} \sum_{j=1}^{N_2}({p}[u_{1_i},u_{2_j}] - \bar{p})^2} \sqrt{\sum_{i=1}^{N_1} \sum_{j=1}^{N_2}({p}[u_{1_i},u_{2_j}] - \bar{\hat{p}})^2}}}.
\end{equation}
}

\vspace{-5pt} 
\subsection{Results}

\vspace{-5pt} 

The average LSD, NMSE, and NCC across $D$ subjects and both ears for the different approaches are shown in Table~\ref{tab:metrics}. Overall, the proposed \ours{} outperforms the other methods across all metrics, demonstrating the effectiveness of the approach.
 \ours{} effectively incorporates statistical priors from only a few realizations thanks to the wavelet scattering transform, offering a practical approach to maximize the use of the available prior information.
Moreover, it is worth noting that using a simple MLP for unsupervised, one-shot continuous field representation can achieve surprisingly good interpolation. This suggests that, even in the absence of training data, a basic MLP can outperform the widely used conventional SH approach.


\begin{table}[h!]
\vspace{-6pt} 
 \caption{Average error metrics: LSD, NMSE, and NCC across $D$ subjects and both ears for the different approaches. }
\vspace{-10pt} 
 \begin{center}
 \begin{tabular}{l|c|c|c|c}
  {} & SH & NF &  SNF & \ours \\ \hline
  $\downarrow$ LSD  &  6.64 &  6.10  & 6.37  & \textbf{5.34} \\
  $\downarrow$ NMSE  & 0.23 & 0.20  & 0.21 &\textbf{0.14} \\
  $\uparrow$ NCC  & 69.66\% & 84.74 \%   & 79.37\% & \textbf{87.79\%}\\
  \multicolumn{5}{l}{$^*$Values marked in bold are the best performances.} \\
 \end{tabular}
\end{center}
 \label{tab:metrics}
 \vspace{-10pt} 
\end{table}

Figure~\ref{fig: example} shows an example at a high frequency of \SI{14470}{\hertz}. 
Qualitatively, \ours{} closely resembles the ground truth. Its reconstruction maintains the overall structure of NN with more refined details. However, without masking, this refinement can deviate significantly from the NN reconstruction and instead become more similar to the known realization of subject 11 (Fig.~\ref{fig: 1g}). This highlights the necessity of masking the scattering coefficients; without it, the method risks merely replicating the known realization. 
Additionally, it is evident that the SH reconstruction introduces some artifacts and fails to  capture the lobe structure, whereas the NN is able to do so.

 \vspace{-5pt} 
\section{Conclusion}
\vspace{-3pt} 
\label{sec:conclusion}
In this paper, we propose the MSNF framework that leverages the WST as a multi-scale feature extractor to impose statistical priors under sparse observation conditions. 
Validation results indicate that the framework offers a promising approach for signal reconstruction in scenarios with limited data, thanks to the use of WST. 

\vspace{-7pt}

\bibliographystyle{IEEEtran}
\bibliography{refs}

@inproceedings{zhang2023hrtf,
  title={HRTF field: Unifying measured HRTF magnitude representation with neural fields},
  author={Zhang, You and Wang, Yuxiang and Duan, Zhiyao},
  booktitle = {Proc. IEEE Int. Conf. Acoust., Speech Signal Process. (ICASSP)},
  pages={1--5},
  year={2023},
  organization={IEEE}
}

@inproceedings{angles2018generative,
  title={Generative networks as inverse problems with Scattering transforms},
  author={Angles, Tom{\'a}s and Mallat, St{\'e}phane},
  booktitle = {Proc. Int. Conf. Learn. Represent. (ICLR)},
  year={2018}
}

@inproceedings{lostanlen2019shape,
  title={The Shape of RemiXXXes to Come: Audio texture synthesis with time--frequency scattering},
  author={Lostanlen, Vincent and Hecker, Florian},
booktitle = {Proc. Int. Conf. Digital Audio Effects (DAFx)},
  year={2019}
}

@inproceedings{masuyama2025retrieval,
  title={Retrieval-augmented neural field for HRTF upsampling and personalization},
  author={Masuyama, Yoshiki and Wichern, Gordon and Germain, Fran{\c{c}}ois G and Ick, Christopher and Le Roux, Jonathan},
booktitle = {Proc. IEEE Int. Conf. Acoust., Speech Signal Process. (ICASSP)},
  pages={1--5},
  year={2025},
  organization={IEEE}
}

@article{wang2022adaptive,
  title={Adaptive scattering transforms for playing technique recognition},
  author={Wang, Changhong and Benetos, Emmanouil and Lostanlen, Vincent and Chew, Elaine},
journal = {IEEE/ACM Trans. Audio Speech Lang. Process.},
  volume={30},
  pages={1407--1421},
  year={2022},
  publisher={IEEE}
}

@ARTICLE{han2024learning,
  author={Han, Han and Lostanlen, Vincent and Lagrange, Mathieu},
  journal = {IEEE/ACM Trans. Audio Speech Lang. Process.},
  title={Learning to Solve Inverse Problems for Perceptual Sound Matching}, 
  year={2024},
  volume={32},
  number={},
  pages={2605-2615},
  doi={10.1109/TASLP.2024.3393738}}

@article{masuyama2025sudafield,
  title={SuDaField: Subject-and Dataset-Aware Neural Field for HRTF Modeling},
author = {{Y. Masuyama} and {G. Wichern} and {F. G. Germain} and {C. Ick} and {J. Le Roux}},
journal = {IEEE Open J. Signal Process.},
  volume={6},
  pages={1169--1178},
  year={2025},
  publisher={IEEE}
}

@article{hogg2025listener,
  title={Listener acoustic personalisation challenge-LAP24: Head-related transfer function upsampling},
  author={Hogg, Aidan OT and Barumerli, Roberto and Daugintis, Rapolas and Poole, Katarina C and Brinkmann, Fabian and Picinali, Lorenzo and Geronazzo, Michele},
journal = {IEEE Open J. Signal Process.},
  year={2025},
  publisher={IEEE}
}

@article{verburg2025differentiable,
  title={Differentiable physics for sound field reconstruction},
  author={Verburg, Samuel A and Fernandez-Grande, Efren and Gerstoft, Peter},
  journal = {J. Acoust. Soc. Am.},
  volume={158},
  number={5},
  pages={4059--4069},
  year={2025},
  publisher={AIP Publishing}
}

@article{fernandez2023generative,
  title={Generative models for sound field reconstruction},
  author={Fernandez-Grande, Efren and Karakonstantis, Xenofon and Caviedes-Nozal, Diego and Gerstoft, Peter},
  journal = {J. Acoust. Soc. Am.},
  volume={153},
  number={2},
  pages={1179--1190},
  year={2023},
  publisher={AIP Publishing}
}

@article{olivieri2024physics,
  title={Physics-informed neural network for volumetric sound field reconstruction of speech signals},
  author={Olivieri, Marco and Karakonstantis, Xenofon and Pezzoli, Mirco and Antonacci, Fabio and Sarti, Augusto and Fernandez-Grande, Efren},
journal = {EURASIP J. Audio Speech Music Process.},
  volume={2024},
  number={1},
  pages={42},
  year={2024},
  publisher={Springer}
}

@article{olivieri2021physics,
  title={A physics-informed neural network approach for nearfield acoustic holography},
  author={Olivieri, Marco and Pezzoli, Mirco and Antonacci, Fabio and Sarti, Augusto},
  journal={Sensors},
  volume={21},
  number={23},
  pages={7834},
  year={2021},
  publisher={MDPI}
}

@article{luan2025pisfd,
  author={Luan, Xinmeng and Pezzoli, Mirco and Antonacci, Fabio and Sarti, Augusto},
  journal = {IEEE/ACM Trans. Audio Speech Lang. Process.},
  title={Physics-Informed Neural Network-Driven Sparse Field Discretization Method for Near-Field Acoustic Holography}, 
  year={2025},
  volume={33},
  number={},
  pages={4282-4294},
  doi={10.1109/TASLPRO.2025.3619768}}

@inproceedings{anden2015joint,
  title={Joint time-frequency scattering for audio classification},
author = {{J. And\'en} and {V. Lostanlen} and {S. Mallat}},
booktitle = {Proc. IEEE Int. Workshop Mach. Learn. Signal Process. (MLSP)},
  pages={1--6},
  year={2015},
  organization={IEEE}
}

@article{albeverio1975homogeneous,
  title={Homogeneous random fields and statistical mechanics},
  author={Albeverio, Sergio and H{\o}egh-Krohn, Raphael},
journal = {J. Funct. Anal.},
  volume={19},
  number={3},
  pages={242--272},
  year={1975},
  publisher={Elsevier}
}

@inproceedings{vyassoap,
  title={SOAP: Improving and Stabilizing Shampoo using Adam for Language Modeling},
  author={Vyas, Nikhil and Morwani, Depen and Zhao, Rosie and Shapira, Itai and Brandfonbrener, David and Janson, Lucas and Kakade, Sham M},
 booktitle = {Proc. Int. Conf. Learn. Represent. (ICLR)},
  year={2025},
}

@article{morel2025scale,
  title={Scale dependencies and self-similar models with wavelet scattering spectra},
  author={Morel, Rudy and Rochette, Gaspar and Leonarduzzi, Roberto and Bouchaud, Jean-Philippe and Mallat, St{\'e}phane},
  journal={Applied and Computational Harmonic Analysis},
  volume={75},
  pages={101724},
  year={2025},
  publisher={Elsevier}
}

@article{cheng2024scattering,
  title={Scattering spectra models for physics},
  author={Cheng, Sihao and Morel, Rudy and Allys, Erwan and M{\'e}nard, Brice and Mallat, St{\'e}phane},
  journal={PNAS nexus},
  volume={3},
  number={4},
  pages={pgae103},
  year={2024},
  publisher={Oxford University Press US}
}

@article{mousset2024generative,
  title={Generative models of astrophysical fields with scattering transforms on the sphere},
  author={Mousset, Louise and Allys, Erwan and Price, Matthew A and Aumont, Jonathan and Delouis, J-M and Montier, Ludovic and McEwen, Jason D},
  journal={Astronomy \& Astrophysics},
  volume={691},
  pages={A269},
  year={2024},
  publisher={EDP Sciences}
}

@article{delouis2022non,
  title={Non-Gaussian modelling and statistical denoising of Planck dust polarisation full-sky maps using scattering transforms},
  author={Delouis, J-M and Allys, Erwan and Gauvrit, Edouard and Boulanger, Fran{\c{c}}ois},
  journal={Astronomy \& Astrophysics},
  volume={668},
  pages={A122},
  year={2022},
  publisher={EDP Sciences}
}

@article{ma2023spatial,
  title={Spatial upsampling of head-related transfer functions using a physics-informed neural network},
  author={Ma, Fei and Abhayapala, Thushara D and Samarasinghe, Prasanga N and Chen, Xingyu},
  journal={arXiv preprint arXiv:2307.14650},
  year={2023}
}

@article{bruna2013invariant,
  title={Invariant scattering convolution networks},
  author={Bruna, Joan and Mallat, St{\'e}phane},
journal = {IEEE Trans. Pattern Anal. Mach. Intell.},
  volume={35},
  number={8},
  pages={1872--1886},
  year={2013},
  publisher={IEEE}
}

@article{oyallon2018scattering,
  title={Scattering networks for hybrid representation learning},
  author={Oyallon, Edouard and Zagoruyko, Sergey and Huang, Gabriel and Komodakis, Nikos and Lacoste-Julien, Simon and Blaschko, Matthew and Belilovsky, Eugene},
  journal = {IEEE Trans. Pattern Anal. Mach. Intell.},
  volume={41},
  number={9},
  pages={2208--2221},
  year={2018},
  publisher={IEEE}
}

@article{mallat2012group,
  title={Group invariant scattering},
  author={Mallat, St{\'e}phane},
journal = {Commun. Pure Appl. Math.},
  volume={65},
  number={10},
  pages={1331--1398},
  year={2012},
  publisher={Wiley Online Library}
}

@article{bruna2019multiscale,
  title={Multiscale sparse microcanonical models},
  author={Bruna, Joan and Mallat, St{\'e}phane},
journal = {Math. Stat. Learn.},
  volume={1},
  number={3},
  pages={257--315},
  year={2019}
}

@article{tancik2020fourier,
  title={Fourier features let networks learn high frequency functions in low dimensional domains},
  author={Tancik, Matthew and Srinivasan, Pratul and Mildenhall, Ben and Fridovich-Keil, Sara and Raghavan, Nithin and Singhal, Utkarsh and Ramamoorthi, Ravi and Barron, Jonathan and Ng, Ren},
  journal={Adv. Neural Inf. Process. Syst. (NeurIPS)},
  volume={33},
  pages={7537--7547},
  year={2020}
}

@inproceedings{rahaman2019spectral,
  title={On the spectral bias of neural networks},
  author={Rahaman, Nasim and Baratin, Aristide and Arpit, Devansh and Draxler, Felix and Lin, Min and Hamprecht, Fred and Bengio, Yoshua and Courville, Aaron},
  booktitle={Int. Conf. Mach. Learn. (ICML)},
  pages={5301--5310},
  year={2019},
  organization={PMLR}
}

@article{brinkmann2019cross,
  title={A cross-evaluated database of measured and simulated HRTFs including 3D head meshes, anthropometric features, and headphone impulse responses},
  author={Brinkmann, Fabian and Dinakaran, Manoj and Pelzer, Robert and Grosche, Peter and Voss, Daniel and Weinzierl, Stefan},
journal = {J. Audio Eng. Soc.},
  volume={67},
  number={9},
  pages={705--718},
  year={2019},
  publisher={Audio Engineering Society}
}

@inproceedings{chen2018gradnorm,
  title={Gradnorm: Gradient normalization for adaptive loss balancing in deep multitask networks},
  author={Chen, Zhao and Badrinarayanan, Vijay and Lee, Chen-Yu and Rabinovich, Andrew},
booktitle = {Proc. Int. Conf. Mach. Learn. (ICML)},
  pages={794--803},
  year={2018},
  organization={PMLR}
}

@article{andreux2020kymatio,
  title={Kymatio: Scattering transforms in python},
  author={Andreux, Mathieu and Angles, Tom{\'a}s and Exarchakis, Georgios and Leonarduzzi, Roberto and Rochette, Gaspar and Thiry, Louis and Zarka, John and Mallat, St{\'e}phane and And{\'e}n, Joakim and Belilovsky, Eugene and others},
journal = {J. Mach. Learn. Res.},
  volume={21},
  number={60},
  pages={1--6},
  year={2020}
}

@article{ahrens2020computation,
  title={Computation of spherical harmonic representations of source directivity based on the finite-distance signature},
  author={Ahrens, Jens and Bilbao, Stefan},
journal = {IEEE/ACM Trans. Audio Speech Lang. Process.},
  volume={29},
  pages={83--92},
  year={2020},
  publisher={IEEE}
}

@inproceedings{zotkin2009regularized,
  title={Regularized HRTF fitting using spherical harmonics},
  author={Zotkin, Dmitry N and Duraiswami, Ramani and Gumerov, Nail A},
booktitle = {Proc. IEEE Workshop Appl. Signal Process. Audio Acoust. (WASPAA)},
  pages={257--260},
  year={2009},
  organization={IEEE}
}

@book{williams1999fourier,
  title={Fourier acoustics: sound radiation and nearfield acoustical holography},
  author={Williams, Earl G},
  year={1999},
  publisher={Elsevier}
}


\end{document}